\documentclass[preprint]{elsarticle}
\usepackage{graphicx}
\usepackage{epsfig}
\usepackage{epstopdf}
\usepackage{float}
\usepackage[usenames]{color}
\usepackage{color,epsfig,amsmath,amssymb,fancyhdr}   
\usepackage{lineno}

\journal{NIM A}

\begin{document}

\begin{frontmatter}

\title{Composite mirror facets for ground based gamma ray astronomy}

\fntext[cor]{Corresponding author, email: cmedina@cea.fr}

\author{P.~Brun}
\author{P-H.~Carton}
\author{D.~Durand}
\author{J-F.~Glicenstein}
\author{C.~Jeanney} 
\author{M.C.~Medina\fnref{cor}} 
\author{P.~Micolon}
\author{B.~Peyaud}
\address{CEA, Irfu, Centre de Saclay, F-91191 Gif sur Yvette, France}

\date{\today}

\begin{abstract}

Composite mirrors for gamma-ray astronomy have been developed to fulfill the specifications required for the next generation of Cherenkov telescopes represented by CTA (Cherenkov Telescope Array). In addition to the basic requirements on focus and reflection efficiency, the mirrors have to be stiff, lightweight, durable and cost efficient. In this paper, the technology developed to produce such mirrors is described,  as well as some tests that have been performed to validate them. It is shown that these mirrors comply with the needs of CTA, making them good candidates for use on a significant part of the array.

\end{abstract}

\begin{keyword}

Atmospheric Cherenkov telescopes \sep mirrors \sep CTA.

\end{keyword}

\end{frontmatter}

\section{Introduction}

Gamma-ray telescopes are built to image $\gamma$-ray induced particle showers in the atmosphere. The telescopes are deployed in arrays in order to obtain a stereoscopic view of the atmospheric event, which allow efficient off-line background subtraction and a better determination of the $\gamma$-ray arrival direction. Moreover, to gain significantly in sensitivity, the next generation of Cherenkov telescopes requires a very large reflective surface area compared with current instruments. See Tab.~\ref{Tab:mirrors1}. 

The next generation of VHE (Very High Energy) $\gamma$-ray telescope array is CTA (Cherenkov Telescope Array), which is currently in the development phase~\cite{2011ExA....32..193A}. Two sites, one in the North and one in the South are planned to provide full-sky coverage. In each of these sites an array of telescopes of multiple sizes will be installed; at minimum there will be small ($\sim$5 m), medium ($\sim$12 m) and large ($\sim$20 m) diameter telescopes (called from now on, Small Size Telescope or SST, Middle Size Telescope or MST and Large Size Telescope or LST, respectively), each optimized for different energy ranges. The final configurations of these arrays are not yet completely defined but  the southern site of CTA will be composed of at least 50 telescopes of 3 different sizes and a total of $\sim$5,000 $\rm m^2$ of mirrors will be necessary.  The northern site, which is intended to be smaller, will require of the order of 3,500 $\rm m^2$ of mirrors. Such a massive production of mirrors has never been conducted for Cherenkov telescopes so far. The mirror area used by the currently running observatories is shown in Tab.~\ref{Tab:mirrors1} .

\begin{table}[htdp]
\caption{Currently running observatories mirrors characteristics.}
\begin{center}
\begin{tabular}{|c|c|c|}
\hline
Instrument & Mirror area [m$^2$]& Mirror type\\ \hline
H.E.S.S. I and II \footnote{www.mpi-hd.mpg.de/hfm/HESS} & 1046 & solid glass\\ \hline
VERITAS\footnote{veritas.sao.arizona.edu} &  440 & solid glass\\ \hline
MAGIC\footnote{magic.mppmu.mpg} & 480  & AlMgSi plates - Al honeycomb\\ \hline
\end{tabular}
\end{center}
\label{Tab:mirrors1}
\end{table}%

\begin{figure}
\centering
\includegraphics[width=.4\textwidth]{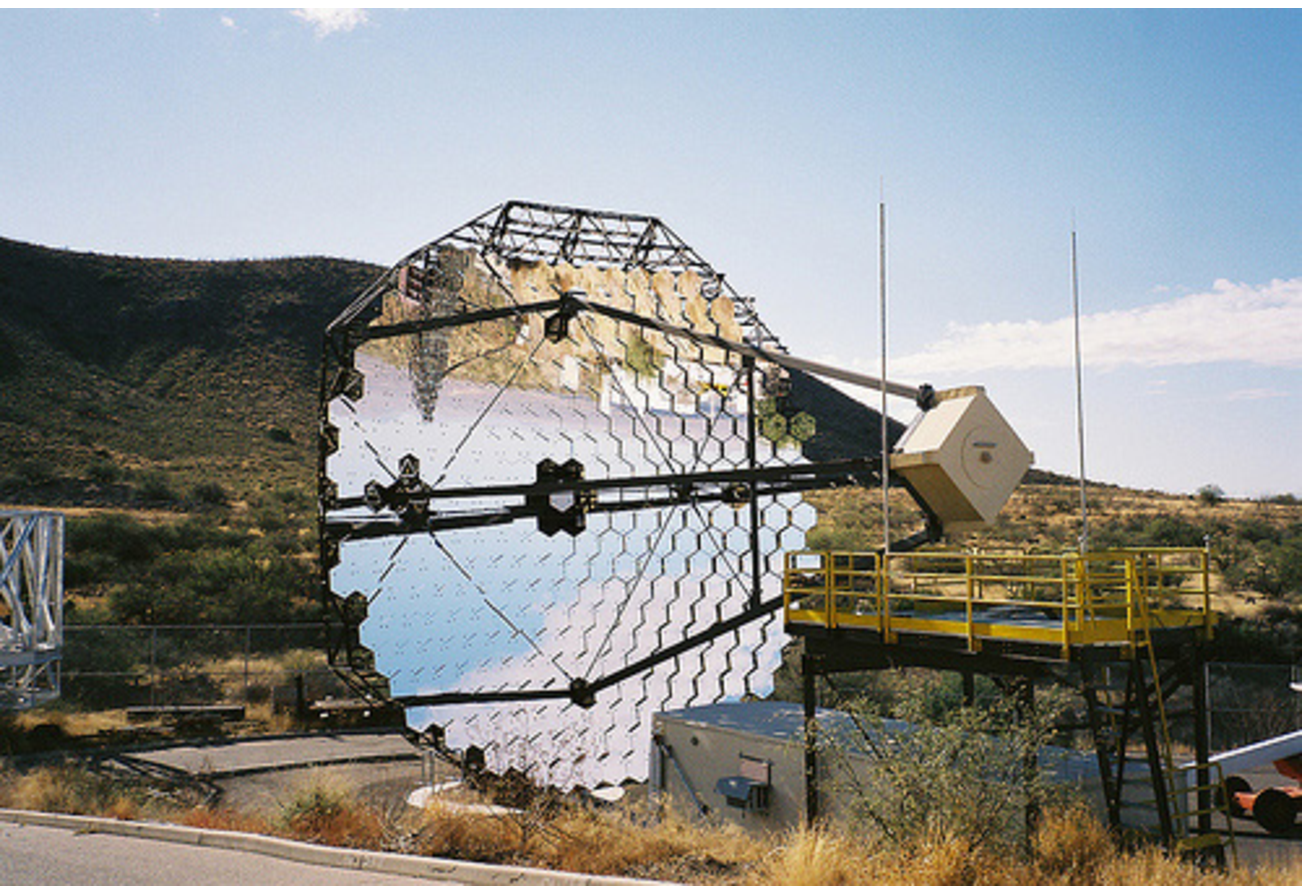}
\includegraphics[width=.35\textwidth]{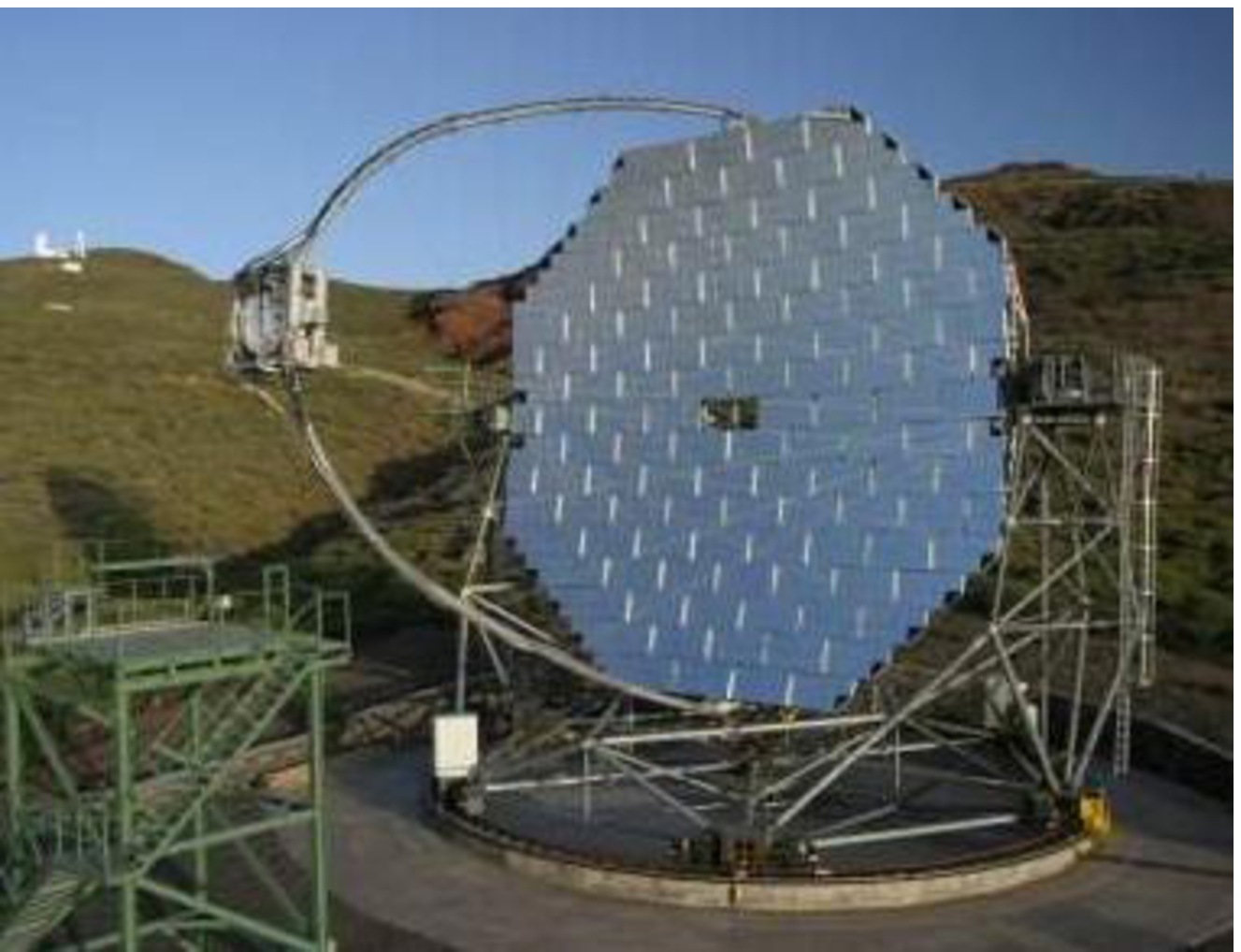}
\caption{Telescopes of the VERITAS array (left) and the MAGIC array (right). The reflectors of these telescopes are composed by small mirror facets. \label{telescopes}}
\end{figure}

To achieve such large dish sizes, the reflector of a Cherenkov telescope is composed of many mirror facets, as shown in Fig.~\ref{telescopes}. For CTA mirrors, the glass solution, as used by H.E.S.S. or VERITAS (see Table~\ref{Tab:mirrors1}) is too costly due to the large surface area to be covered. 
Aside from optical requirements, four aspects drive the choice of mirrors for a large telescope array: 1) cost, 2) weight, which limits the amount of power needed to move the telescopes as well as the construction of the support structure, 3) ease-of-installation, and 4) longevity or durability, which limits the number of times the mirrors need to be resurfaced or replaced within the lifetime of the instrument. 
One solution then, is to develop composite mirrors, as in the MAGIC experiment~\cite{2008SPIE.7018E..28P}.

The mirrors that are described here are developed for the MST. However, this technique could also be suited for Large-Scale Telescopes (LST). MSTs are 12 m diameter telescopes with Davies-Cotton mount~\cite{Davies195716}, meaning that all facets are identical with spherical shape. A telescope would comprise about 100 facets,  all of which should focus Cherenkov light onto the focal plane located at a distance of the order of 16 m. Therefore, the individual facets are segments of a $\sim$30 m diameter reflecting spherical cap. For CTA MST, the baseline idea is to use hexagonal mirrors of 1.2 m diameter (flat to flat), with a spherical shape of about 32 m radius of curvature (precise values are given in the following).

\section{Optical specifications for CTA mirror facets}\label{sec:opt_spec}

The specifications for CTA MST facets rely on the experience of running experiments and Monte-Carlo simulations \cite{2011ExA....32..193A}. The required sensitivity to observe the dim Cherenkov radiation from an extensive air shower is obtained with the use of large reflective area that focuses the light onto a sensitive camera. For a given dish design, the focal distance is fixed  by the desired field of view of a single telescope (mostly driven by the seeked maximum impact parameter for which one wants to detect an atmospheric shower). Once the focal distance is fixed, the pixel size in the focal plane is chosen in order to optimize the sampling of the atmospheric shower image and not to integrate too much night sky background. 
This pixel size defines the PSF of a mirror facet, which must focus light from a point-like source onto a single camera pixel.  
From the work of specifying the telescope parameters conducted within the CTA consortium, the main requirements of  MST mirror facets were obtained: a 12m diameter dish with a focal length of 16.07~m, field of view of $\sim$6$^{\circ}$, and pixel diameter of 5 cm ($\it i.e.$ 3.1 mrad).  In comparison to typical requirements for optical astronomy mirrors, the focusing is about 2 orders of magnitude less strict. This implies in turn a modest tolerance on the distance of the mirror facets to the focal plane. If the focusing capability of the mirrors is good enough, a difference of few centimeters on their focal lengths would be still acceptable. 
The mirrors should have good reflectivity in the wavelength range of 300~nm -- 600~nm, inferred from the spectrum of Cherenkov light after attenuation in the atmosphere.
Each mirror facet should be able to focus at least 85\% of distant parallel incident light into 2/3 of a pixel. Out of this light, 80\% should end up within 1/3 of a pixel. These requirements  translate into the following optical specifications:

\begin{itemize}
\item[-] Focal distance: 16.07 m
\item[-] On-focus reflectance: $\geq$85\% of an incident parallel beam should be reflected into 2 mrad
\item[-] Angular resolution: $\geq$80\% of the above mentioned light should be focused into 1 mrad  
\end{itemize}

\section{Mechanical specifications and durability}

The mirror facets for CTA should be designed to keep the best optical and mechanical performance for more than a decade in operation conditions on site. 

To meet the longevity requirements for CTA instrument, we must define the capacity of the mirror to keep its shape and endurance of the reflective coating. Here, durability refers to the mechanical properties of the mirrors. 
The support structure on the telescope dish is designed with the assumption that mirror facets should weight less than 30 kg, and the thickness of the facets should be 80 mm at most. The stiffness of the mirrors should be such that the optical properties are maintained whatever the orientation of the mirror facet and when submitted to a 50 km/h wind load. They shall also be able to withstand a certain degree of physical impact without a reduction of overall optical quality.

The temperature of operation, inferred from current Cherenkov telescopes sites, would be in the range from -10$^\circ$C to +30$^\circ$C, meaning that the optical performance should be maintained within this range. Moreover, the mirrors should not suffer any irreversible change after being submitted to extreme weather conditions as temperatures in the range from -25$^\circ$C to +60$^\circ$C and a 200 km/h wind load.


\section{Composite mirror technology}
\label{sec:mirror-tech}
\begin{figure}[h]
\centering
\includegraphics[width=.65 \textwidth]{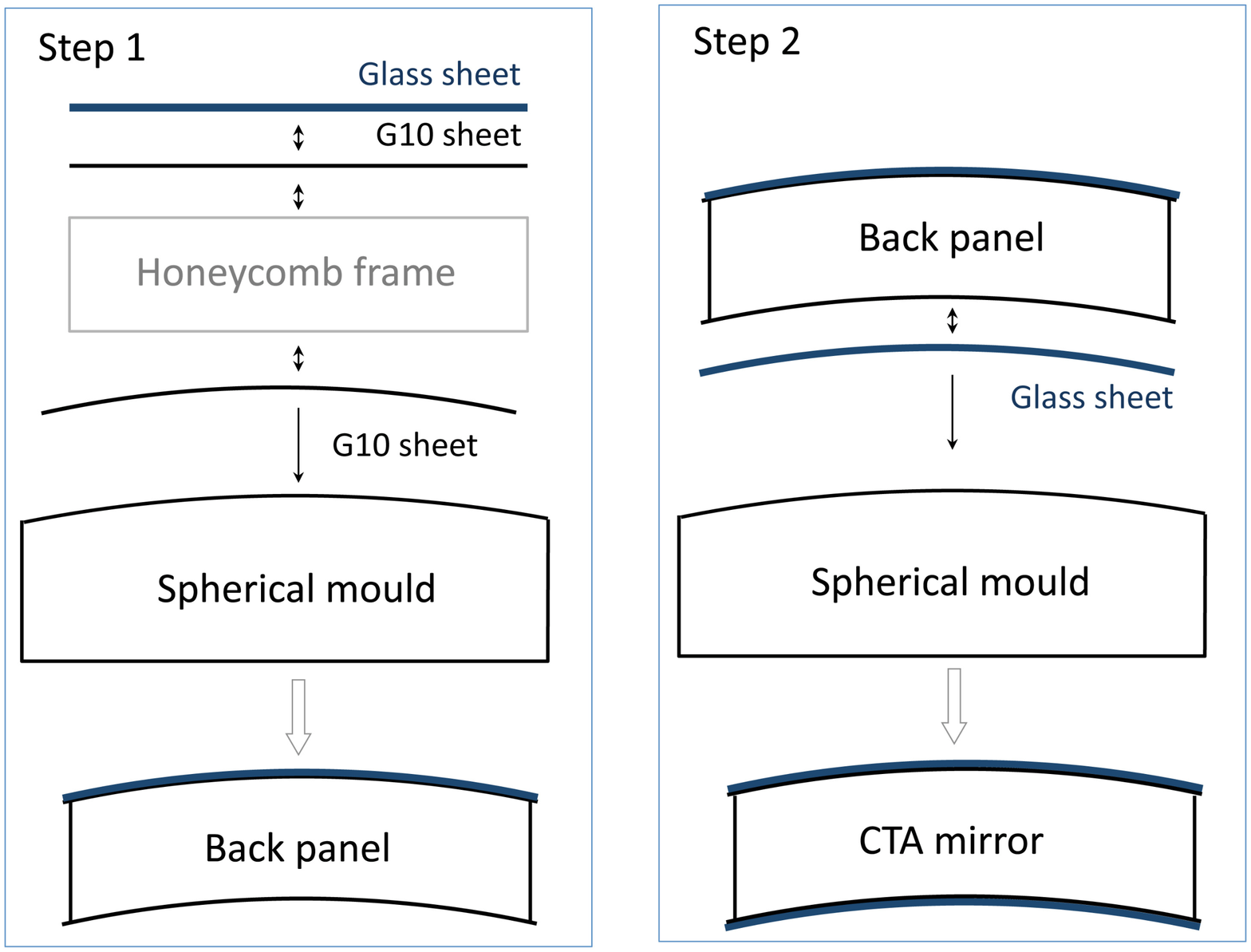}
\caption{\label{steps} Mirror assembly process steps.}
\end{figure}

\begin{figure}[h]
\centering
\includegraphics[width=.4\textwidth]{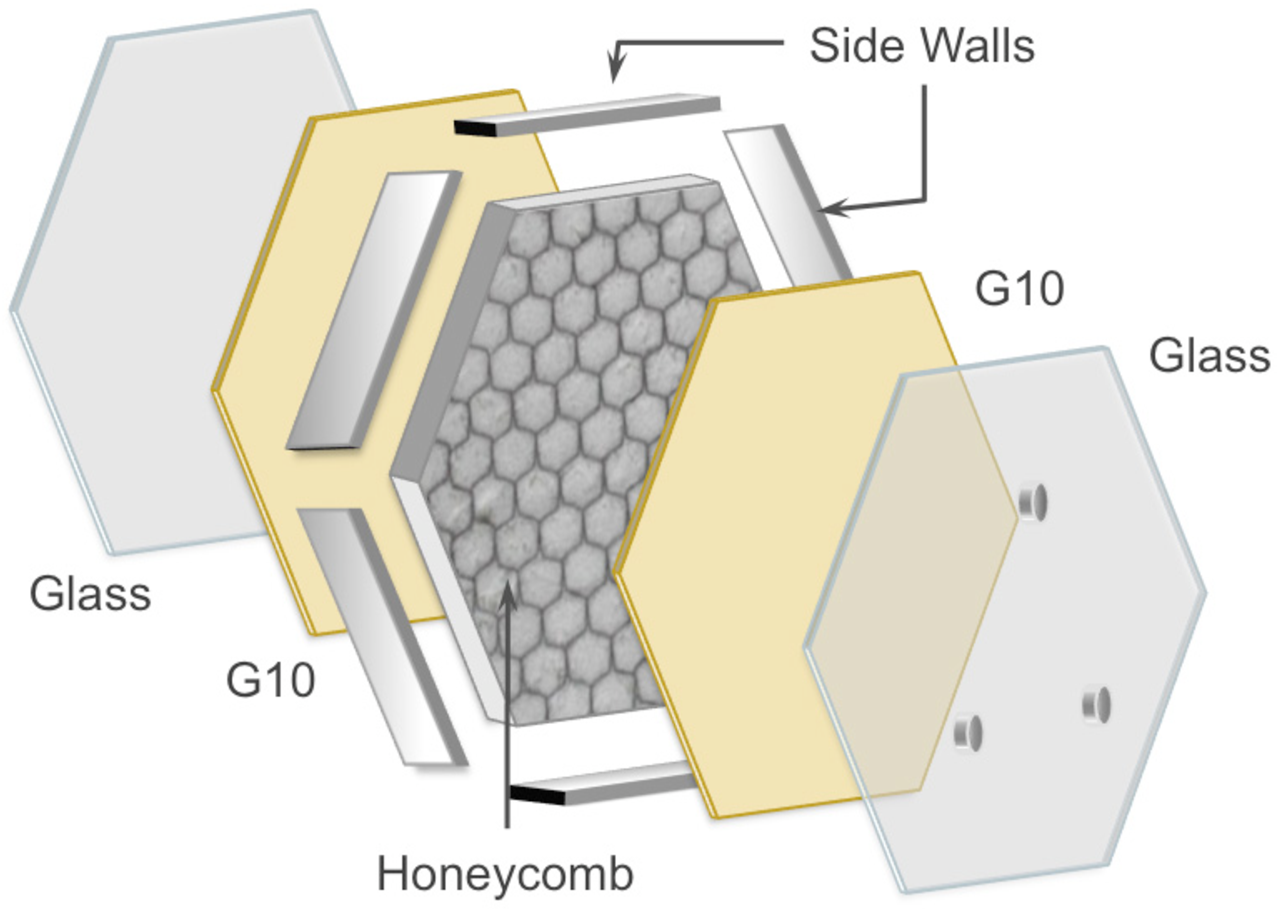}
\includegraphics[width=.4\textwidth]{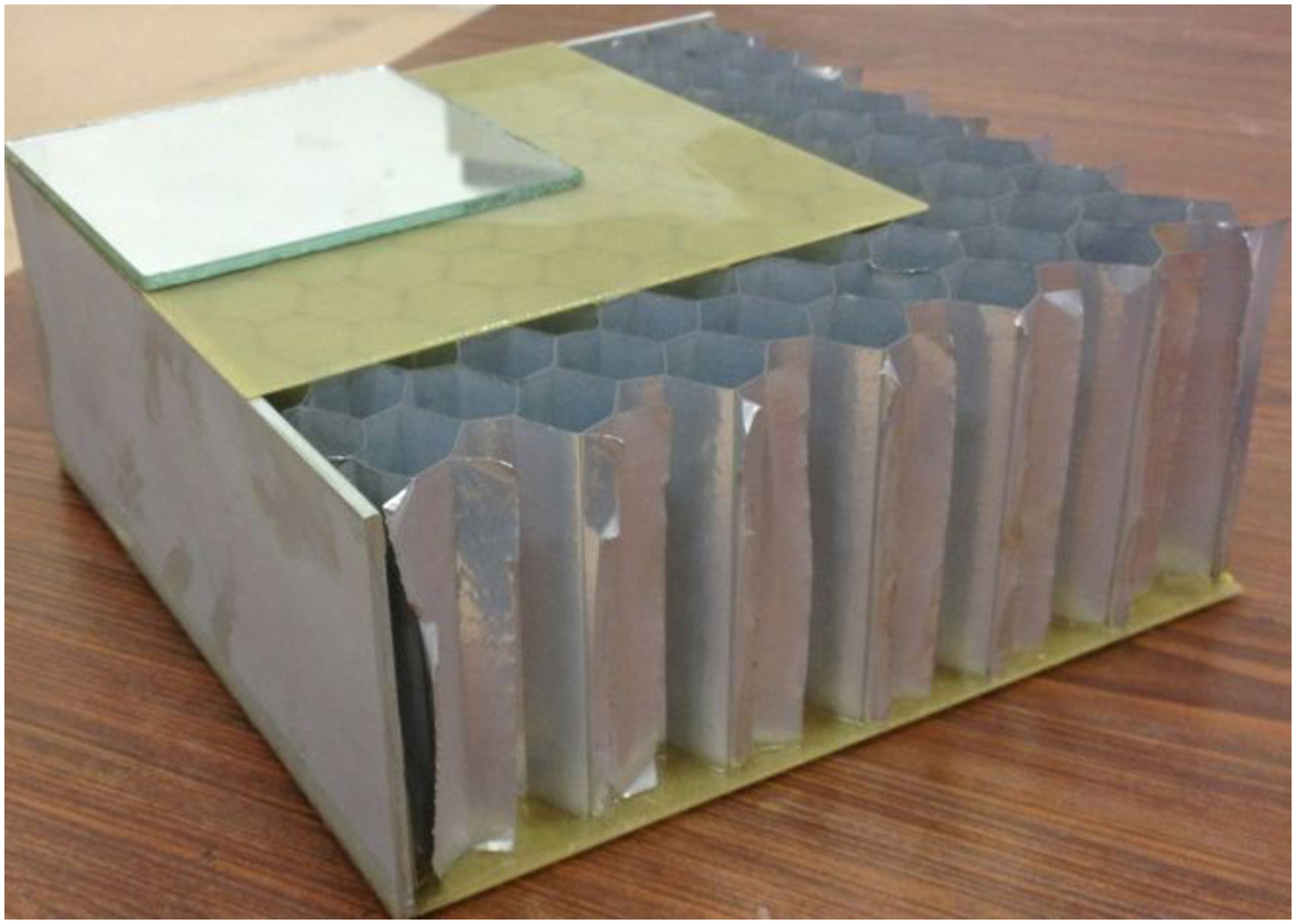}
\caption{\label{expl} Exploded view of the mirrors (left panel), and real exploded model (right panel).}
\end{figure}

The choice for a composite design is dictated by the need to have simultaneously a smooth surface on which to deposit reflecting coating, and a stiff and lightweight body. The idea is to hold a glass sheet that has adequate surface roughness to a support structure made of aluminum honeycomb and glass fiber~\cite{2011arXiv1111.2183C}. The mirrors are produced following a cold slumping technique where the structure is shaped against a spherical mold  at room temperature. The construction is a two-steps process, shown in Fig.~\ref{steps}. In the first step, the back panel is assembled and glued against the mold. The structure obtained is a spherical shell. The RMS deviation from the nominal sphere at this stage is about 12~$\mu$m. As a second step, two glass sheets are glued to each face of the shell. After this stage, the mirror is ready to receive its reflective layer. The gluing of the front glass sheet has the effect of smoothing out the defects, so that the RMS deviation from the nominal sphere after gluing the glass is less than 5~$\mu$m. This value is to be compared to the corresponding value on the mold, which is 7~$\mu$m (the glass sheet somehow flattens out the defects of the mold). In the process, thick aluminum walls are added to each of the six sides. This helps constraining the edges of the front surface to bend correctly. Fig.~\ref{expl} shows an exploded view as well as an actual cutaway view of the mirrors. In Fig.~\ref{steps} and Fig.~\ref{expl}, the layers of glue are not displayed. Different types of glue were tested and standard room temperature polymerizing Araldite resin provides the required stiffness and stability.

On the right panel of Fig.~\ref{expl} one can see all components of the mirrors:
\begin{itemize}
\item[-] 2 glass sheets (2 mm)
\item[-] 2 glass fiber (G10) sheets (2 mm)
\item[-] The aluminum honeycomb, 80 mm height, with hexagonal cells of 19 mm diameter
\item[-] Aluminum side walls (3 mm).
\end{itemize}

The aluminum foils of the honeycomb are 50 $\rm \mu m$ thick and are micro-perforated. That is important when the mirror is held against the mold using vacuum suction. Thanks to these micro holes, the pressure when the mirror is held against the mold using vacuum suction is identical. The fact that the honeycomb foils are thin (50~$\mu$m) makes the raw structure (before gluing the G10 sheets) rather flexible in spite of the honeycomb thickness. It is then possible to enforce a spherical shape without the need to have it milled.

The fact that the structure is front-back symmetric (made with the same components) is very important for its thermal behavior. Some earlier prototypes built without the rear glass sheet showed strong variations of their effective focal distance with changes of temperature due to the differential thermal dilation between the front and the back. In the baseline design presented here, the fully symmetric structure prevents this behavior, such that no significant change of focal distance has been measured when the ambient temperature changed between 10$^\circ$C and 20$^\circ$C.

The mirrors weight 25 kg and are 85 mm thick. On the rear face, 3 pads are glued on a 640 mm radius circle for the mirrors to be held through the standard CTA 3-point supports. At the end of the fabrication process, the mirror is sealed to make it air tight. Even if air is trapped inside the mirror, it has been checked that a difference of pressure up to $\pm$ 50 mbar between the inside of the outside has no influence on its optical properties.

\section{Front face coating}

The coatings of the mirrors have been performed by an industrial partner, Kerdry Thin Film technologies\footnote{www.kerdry.com}. The coating is applied using a standard process where material is vaporized in a large vacuum chamber in which the mirror is affixed face-back at the top and spun. 
At early stage of the mirror development, back-coating has also been considered. Using thin glass sheets allows to minimize the attenuation in the glass and to protect the reflective layer. However, that solution has been abandoned as it has been shown that it enhances the formation of ice and dew on the mirror surface, making it unusable in cold weather periods.

\begin{figure}[!h]
\centering
\includegraphics[width=.4\textwidth]{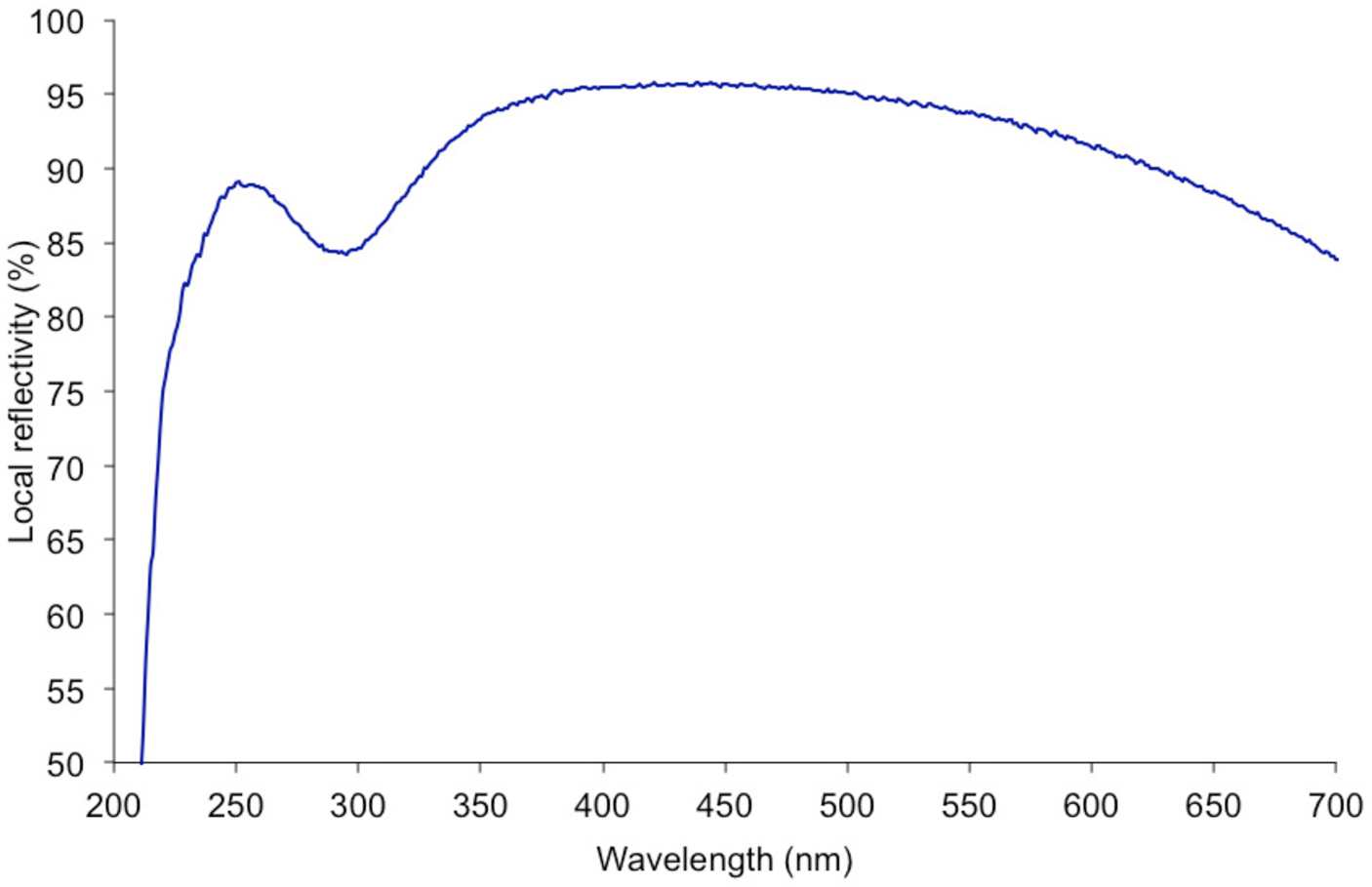}
\includegraphics[width=.4\textwidth]{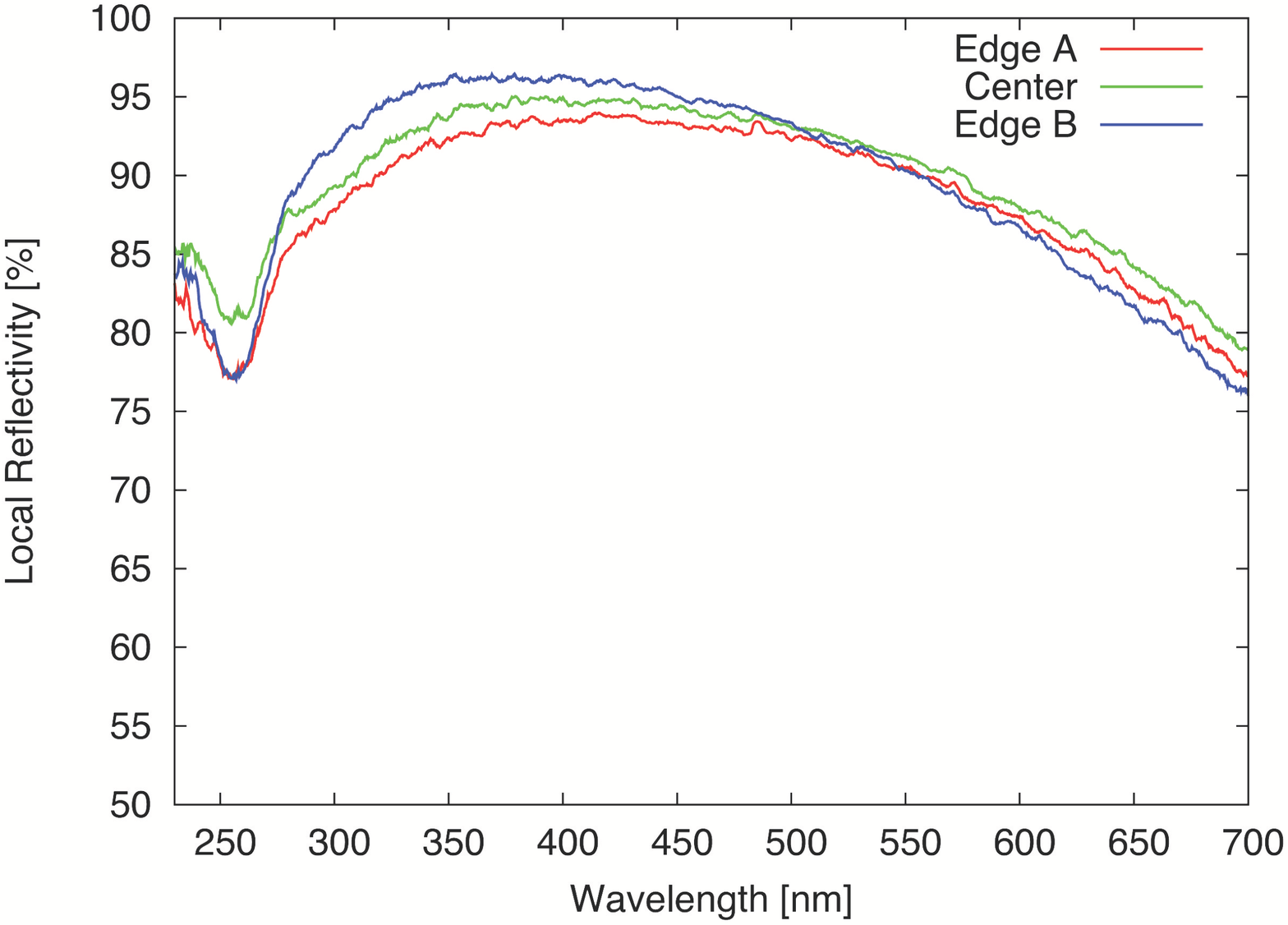}
\caption{\label{reflectivity} {\it Left}: Local reflectivity of the coating between 200 nm and 700 nm measured on a small mirror sample aluminized together with the prototype mirrors.  {\it Right}: Local reflectivity of the coating of the nominal size mirror measured at three different positions. From this it's possible to see that coating properties are uniform within  5~\%.}
\end{figure}

The coating that is used on the presented mirrors is made of an aluminum layer plus three additional layers for protection:  $\rm SiO_2$, $\rm HfO_2$,$\rm SiO_2$. 
Two layers thickness were tested, 120~nm and 240~nm, to determine the effect on durability. 
The locally measured reflectivity of the mirrors is displayed in Fig.~\ref{reflectivity} in the 200~nm to 700~nm wavelength range in the case of 120nm 3 layers protection.

\section{Optical tests}

\subsection{Test bench setup}

In order to characterize the prototype mirrors and evaluate them according to the CTA Consortium optical specifications, a ``2f'' dedicated test bench has been built in an underground room at IRFU\footnote{CEA, Saclay, France.}. This facility allows the measurement of three important parameters of the mirrors: the Point Spread Function (PSF), the overall reflectivity ($\rho$)  and the effective focal length (f). The principle of a ``2f'' setup is the following: the mirror is uniformly illuminated by a light source placed at twice the mirror focal length (2f) and close to its optical axis. The light source should be point-like; in practice it is much smaller than the mirror PSF. The 2f method assumes a uniform illumination of the reflective surface, which has to be controlled. The light reflected by the mirror will ideally produce at 2f an inverted 1:1 scale image of the source. The spread of this image is twice the mirror PSF. 

\begin{figure}[!h]
  \centering
   \includegraphics[width = 0.8\textwidth]{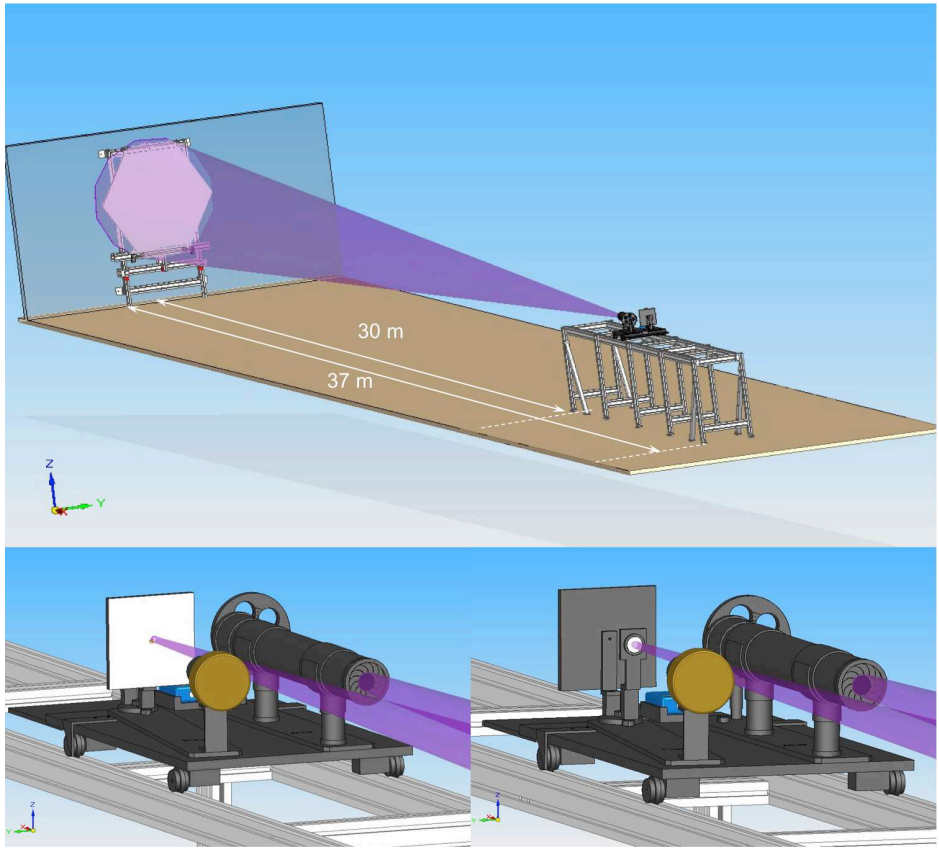}
   \caption{{\it Top}: 2f test bench setup. {\it Bottom, left}: The movable optical table with the screen in place to record the image with the CCD camera. {\it Bottom, right} The screen is replaced by a photodiode when the reflectivity of the mirror is measured.  \label{fig:banc}} 
\end{figure}

\begin{figure}[!h]
  \centering
    \includegraphics[width= 0.5\textwidth]{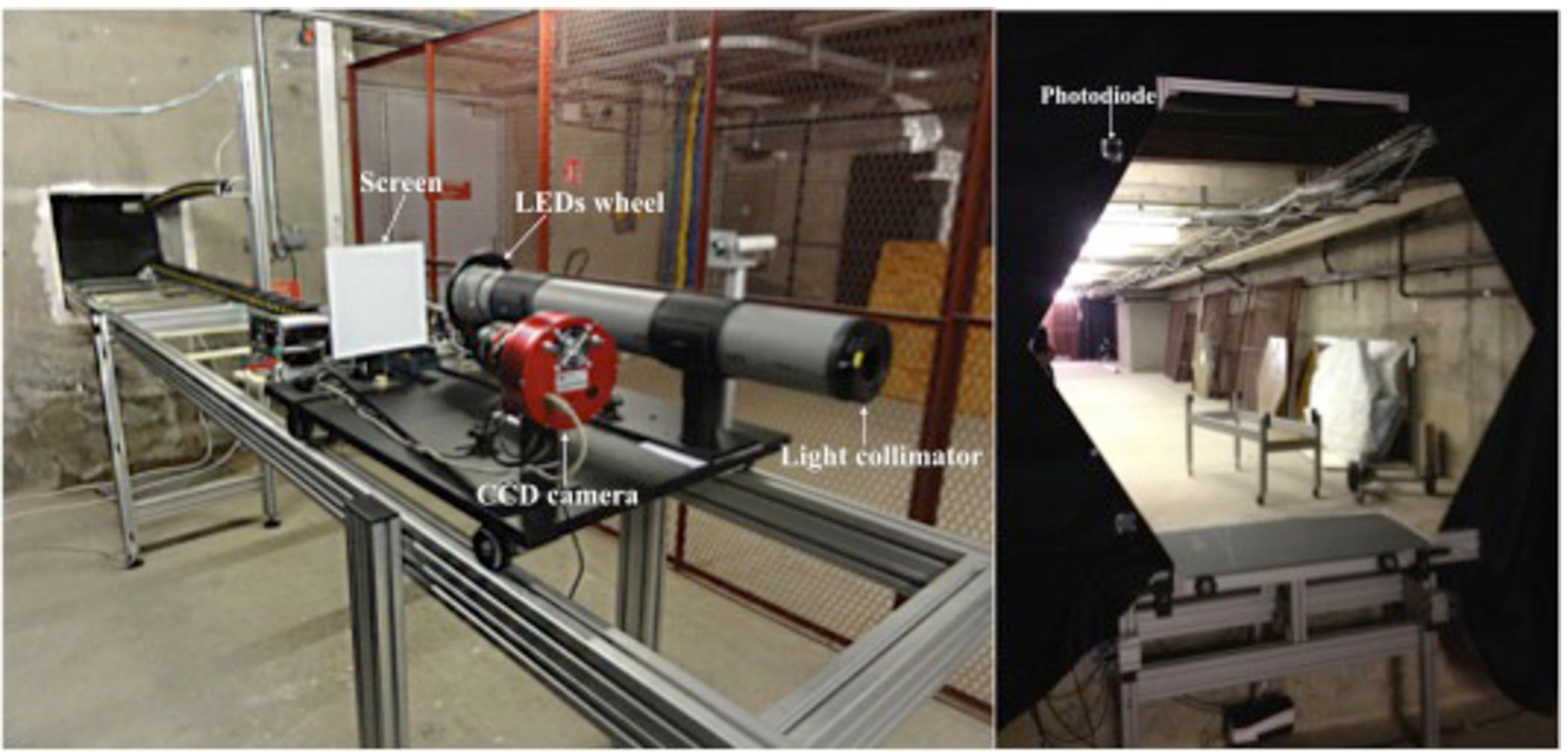}
    \includegraphics[width= 0.184\textwidth]{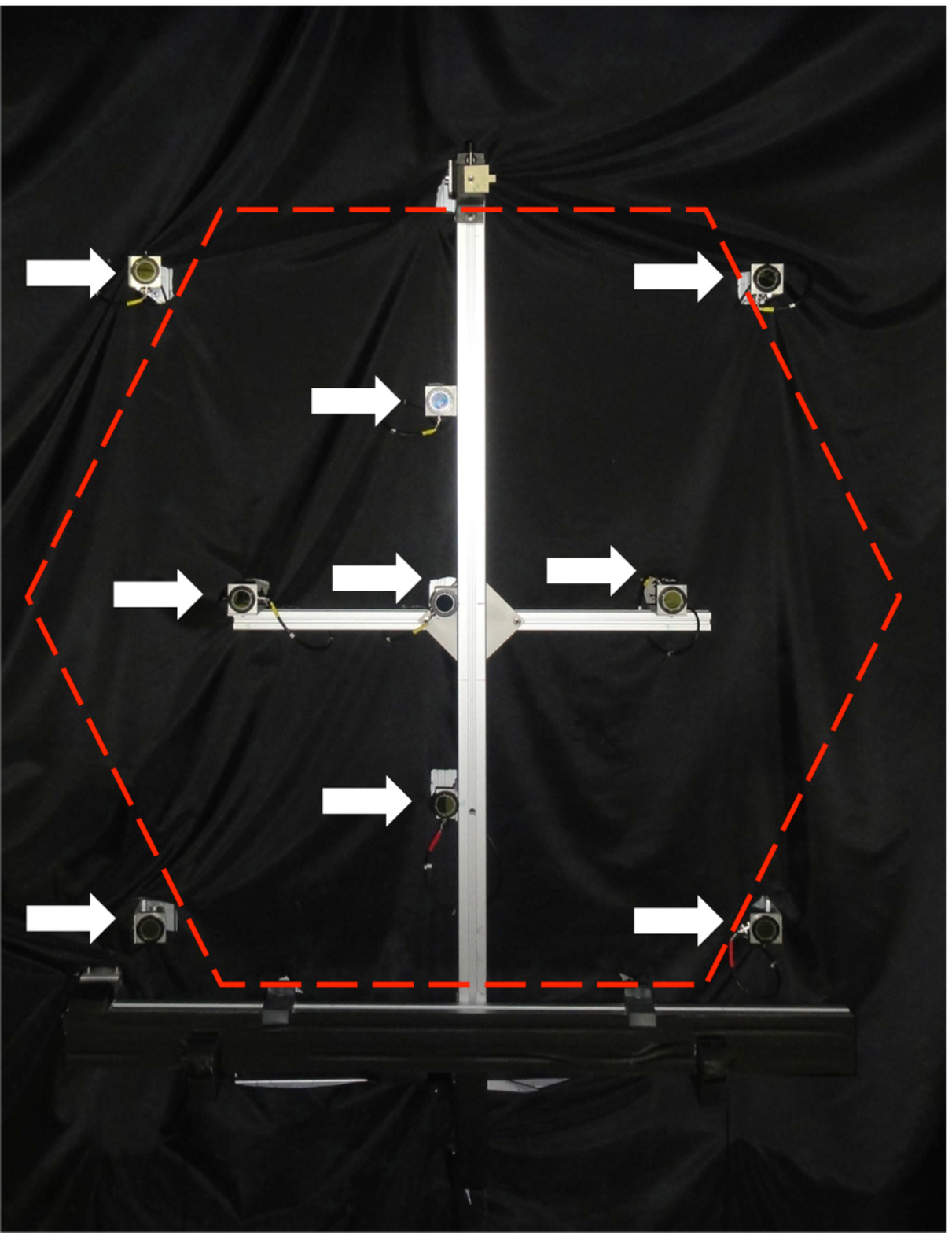}
  \caption{ {\it Left}: Optical table. {\it Center:} Mirror support. {\it Right:} Photodiodes attached to the mirror support (5 of them are behind the mirror for background monitoring and for light pool uniformity measurement.).  \label{fig:banc-1}} 
\end{figure}

This image is formed on a screen placed next to the light source and it is captured by a CCD camera. A standard image reduction procedure is performed, and the images are normalized by the individual pixel response using the flat fielding technique. The electronic noise offset and ambient background are subtracted based on dedicated dark observations. From the analysis of the corrected CCD images (intensity and morphology of the spot) one can derive the PSF of the mirror. This is defined as the diameter of the circle containing 80 $\%$ (``d80'') of the incident light in an circle of 2~mrad diameter (see specifications defined in section~\ref{sec:opt_spec}). 
Fig.~\ref{fig:banc} shows the schematics of the experimental setup. The mirrors are held by a support  attached to the wall. The fine adjustment of the mirror alignment is done remotely in order to place the reflected light spot at the right position on the screen. A movable optical table containing the light sources, the screen, the CCD camera and the photodiode can be displaced along the optical axis allowing a scan of the mirror radius of curvature between 30~m and 37~m. Fig.~\ref{fig:banc-1} shows the actual setup with its principal components.  In order to avoid parasitic reflections, the wall behind the mirror has been covered with a black non-reflective material. The main technical characteristics of the test bench are summarized in Tab.~\ref{table1}.

For the determination of the overall reflectivity of the mirrors, the light flux concentrated at the image position is measured with a large surface photodiode (27.9 mm diameter or 1.9 mrad for MST focal distance) and this value is compared to the total light flux illuminating the mirror. This is obtained by scaling the flux of four similar photodiodes placed at the sides of the mirror to the total surface of the mirror. The crucial point on this determination is the uniformity of the mirror illumination and the stability of the light source during the acquisition. The uniformity of the light pool at the mirror position is checked regularly and it is found to be within $\sim$97\%. The light sources are monitored constantly even though the acquisition time is short enough ($<$~1 min) to avoid any temporal instability. The response of  photodiodes placed near the mirror were normalized with respect to the one at the focusing position for each wavelength. There are two limitations on this measurement: one comes from the low sensitivity of the photodiodes at short wavelengths ($<$400 nm)  and the other from the reduced size of their effective area with respect to the focused spot size defined on the specifications (the photodiode covers only an area of 0.87~mrad, instead of 1~mrad). 
The first point could be solved by using a more intense UV light source and increasing the exposure time. For the second point, a solution could be to use the screen image to derive the global reflectivity. This means a precise characterization of the absolute response of the screen and a cross-calibration between the photodiode and  the CCD response. This could be done in a future upgrade of the test bench. 

\begin{figure}[!h]
  \centering
    \includegraphics[width=0.5\textwidth]{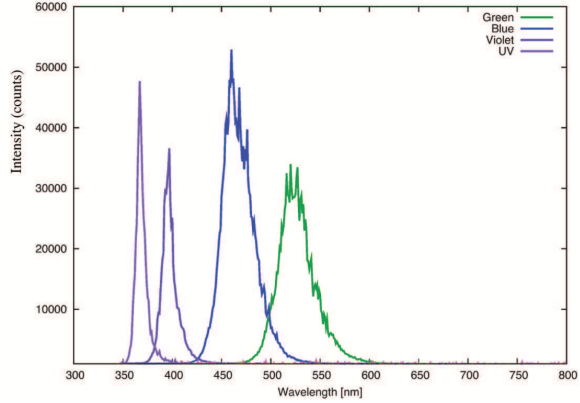}
   \caption{Spectra of the four different LED-type sources used for the optical characterization of the mirrors. \label{fig:spectraLED}} 
\end{figure}

A wheel containing four LED sources is connected to a tube which collimates the light towards the mirror, avoiding spurious light (parasitic reflections, etc) contamination. The four sources have very narrow spectra peaked at 365~nm, 400~nm, 460~nm and 523~nm. The LED emitters are 1.5 mm large, which can be safely considered as point like compared to the PSF of the mirror, which is on the order of 1~cm at twice the focal distance. Fig.~\ref{fig:spectraLED}  shows the four LED spectra measured with a standard spectrometer. 

\begin{table}[!h]
\centering 
	\begin{tabular}{ l  c }
	\multicolumn{2}{c} {2f test bench characteristics}\\ 	\hline \hline 
 	corridor length &  37~m\\ \hline
 	mirror mounting & 3 points support \\ \hline
	alignement of mirrors & remotely controlled \\ \hline
	optical table & movable \\ \hline
 	screen & 20 cm $\times$ 20 cm Alucore (95$\%$ of reflectivity from 400 nm)\\ \hline
  	light source type & LED  + 1.5 mm pinhole \\ \hline
 	CCD camera & ATIK 4000,  sensor Kodak KAI 04022, 1024 $\times$ 1024 pixels \\ \hline
 	photodiodes & Active area of 611 mm$^2$ \\ \hline
	data taking & Windows PC (CCD and photodiodes acquisition)  \\ \hline
 	\end{tabular}
	\caption{Main characteristics of the optical test bench.\label{table1}}
\end{table}

For the determination of the effective focal length, the 80\%-containment diameter (d80) is determined at different distances between the mirror and the plane in
which the light source and the screen are located. The distance at which the d80 is minimal is defined to be the effective focal length of the mirror. The precision on the determination of the distances is of 1~cm.

\subsection{Results: optical performance of prototype mirrors}

\begin{figure}[!h]
  \centering
  \begin{minipage}[L]{7 cm}
    \includegraphics[width=0.98\textwidth]{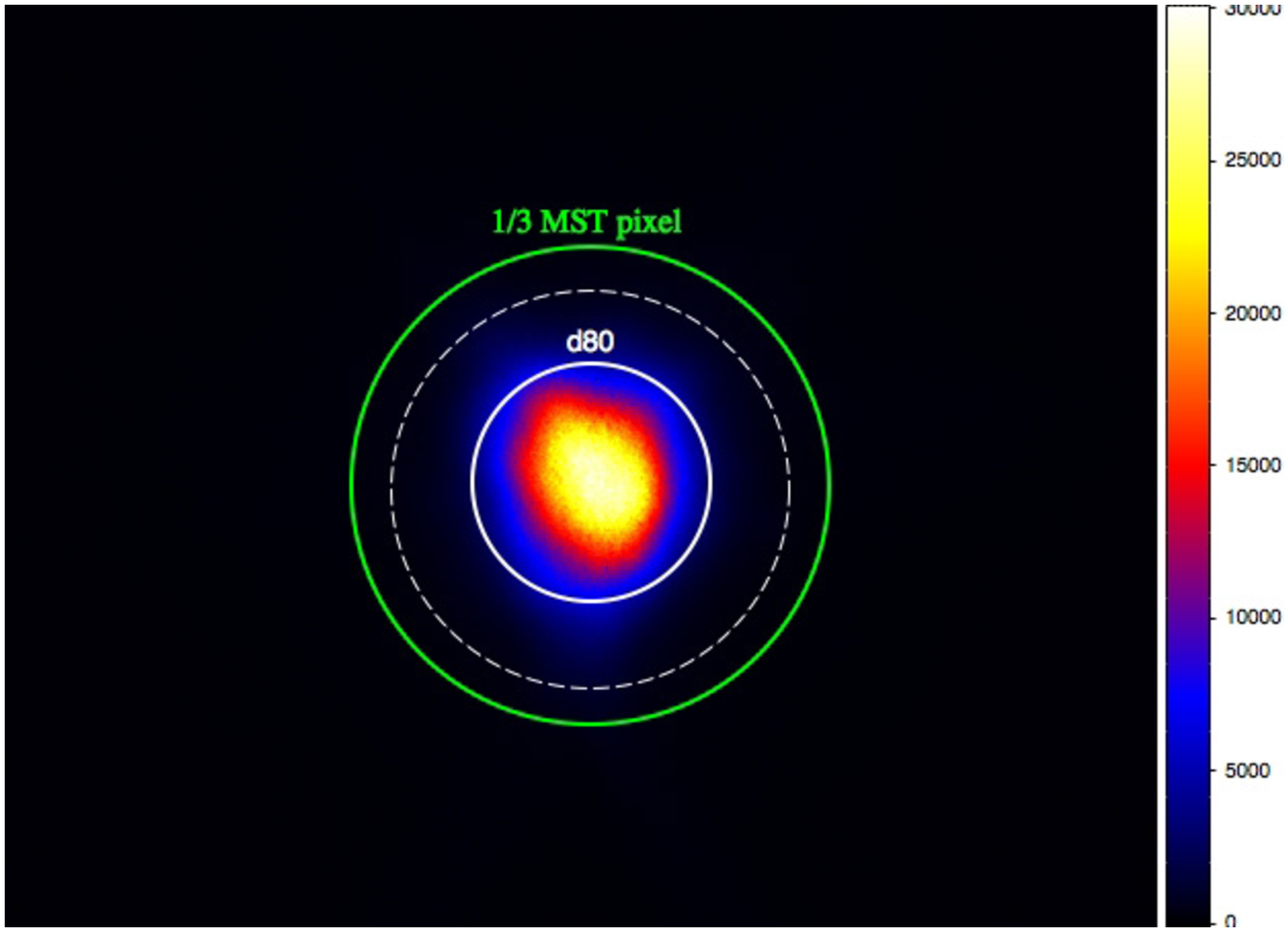}
    \end{minipage}
    \begin{minipage}[R]{7 cm}
    \includegraphics[width=0.89\textwidth]{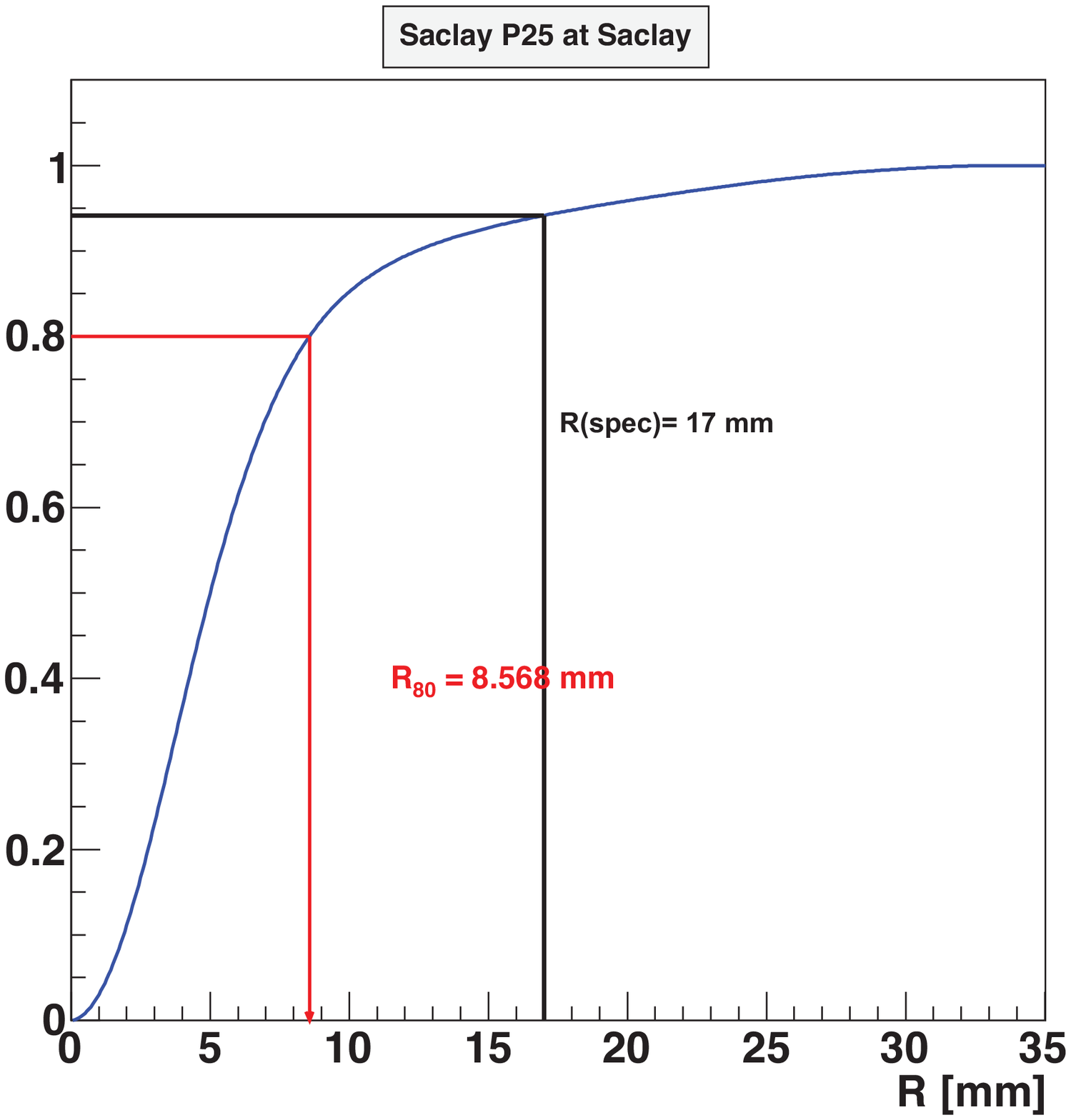}
   \end{minipage}
   \caption{{\it Left}: Image of a LED-type source at twice the nominal focal distance for a prototype mirror. The green circle corresponds to the spot size limit imposed by CTA specification for the MST facets. The white dashed circle represents the surface covered by the photodiode used to measure the focused light flux. The white solid circle represents the 80\% containment diameter d80. {\it Right}: Determination of the radius of the circle containing 80$\%$ of the total light focused on a 2/3 of a MST pixel. \label{fig:PSF}}
\end{figure}

The use of this  test bench was central for evaluating the optical performance of the first pre-series of prototype mirrors. These mirrors were produced using the technique described on section \ref{sec:mirror-tech}. Here we present the results of the most representative prototypes concerning its focusing capability, its global reflectivity and its effective focal length.    
In the left panel of Fig.~\ref{fig:PSF} the CCD image of the spot at the screen produced by the mirror of a light source at 2f is shown. The circle defined by CTA specifications for MST facets (green) and the one corresponding to the photodiode effective area (dashed white) are also shown.  In the right panel of Fig.~\ref{fig:PSF} the radial distribution of the integrated flux from which the d80 is obtained. 

The global reflectivity of this mirror for the four available wavelengths are given in Tab.~\ref{tab:ref}. It is measured to be of the order of 80\% $\pm$ 3\%.  This value fulfills the CTA specifications. In addition one has to notice that the photodiode used here is smaller than the nominal 1~mrad corresponding to 1/3 pixel. It is actually 0.87~mrad diameter at the nominal 2f position. This means that the value obtained with this method represents only a lower limit for the global reflectivity of the mirrors.

The quoted error includes the systematics related to background contamination, detection efficiency, non-uniformity of illumination and dark currents of the photodiodes and the statistical deviation from 10 successive measurements.  This mirror was coated with an Aluminum layer and 3 protective layers (SiO$_2$ + HfO$_2$ + SiO$_2$) with a total thickness of 120 $\mu$m.  

\begin{figure}[!h]
  \centering
    \includegraphics[width=.6\textwidth]{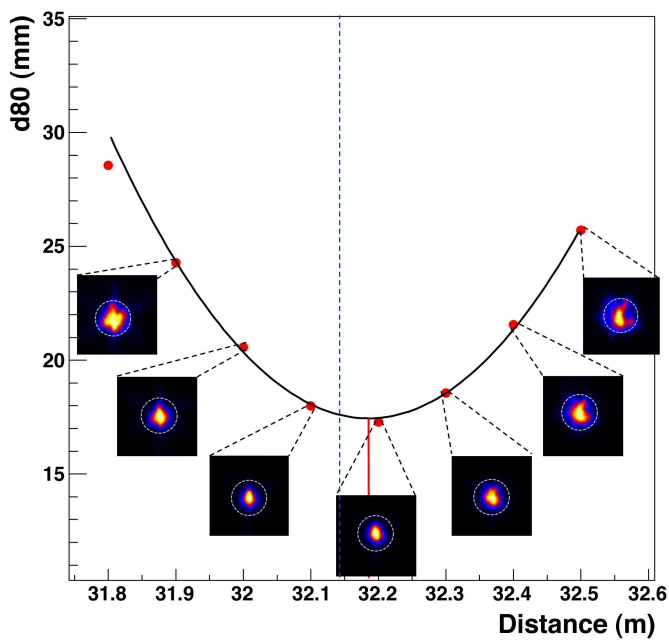}
   \caption{Determination of the effective focal length of the mirror facet. The focused spot size is determined at different distances from the mirror and the minimum corresponds to twice the effective focal length. The minimum of a parabola fitted to the measured d80 at different distances is in agreement with the nominal radius of curvature of the mirror.  The vertical blue dotted line is the nominal radius of curvature. \label{fig:fitFocalDist}}
\end{figure}

Fig.~\ref{fig:fitFocalDist} shows the parabola fit to the spot size, defined as the d80, measured at different distances from the mirror. The minimum of this parabola is positioned at twice the effective focal length of the mirror. Here it can be seen that for this mirror, it corresponds to its nominal value of 32.14 m.

\begin{table}[!h]
  \centering
  \begin{tabular}{|c|c|}  
  	\hline
	Wavelength [nm] & Global Reflectivity [$\%$]  \\ \hline
 	523 & 81  $\pm$ 3 \\ 
	\hline
 	460 & 83 $\pm$ 3 \\ 
	\hline
	400 & 80 $\pm$ 3 \\ 
	\hline
 	365 & 83 $\pm$ 3\\ 
	\hline
	\end{tabular}
	\caption{Measured values of the reflectivity for one prototype mirror for the four wavelengths used.\label{tab:ref}}
\end{table}

\section{Environmental tests}

One of the key issues concerning the composite mirrors presented here is their mechanical stability under quite extreme weather conditions present at many astronomical sites.  The telescopes will not be protected by domes and the mirrors will be continuously in contact with the environment.  Their optical performance should not be degraded after being exposed to large temperature gradients, heavy wind loads, sun irradiation, sandy and salty atmospheres or impacts such as bird beak impacts.   
In order to determine the behavior of the mirrors under such conditions we performed a series of tests, designed to reproduce some of the expected environmental conditions, which are presented in the following sections. 

\subsection{Water tightness and stability test}

\begin{figure}[!h]
  \centering
   \begin{minipage}[R]{7 cm}
    \includegraphics[width=7cm]{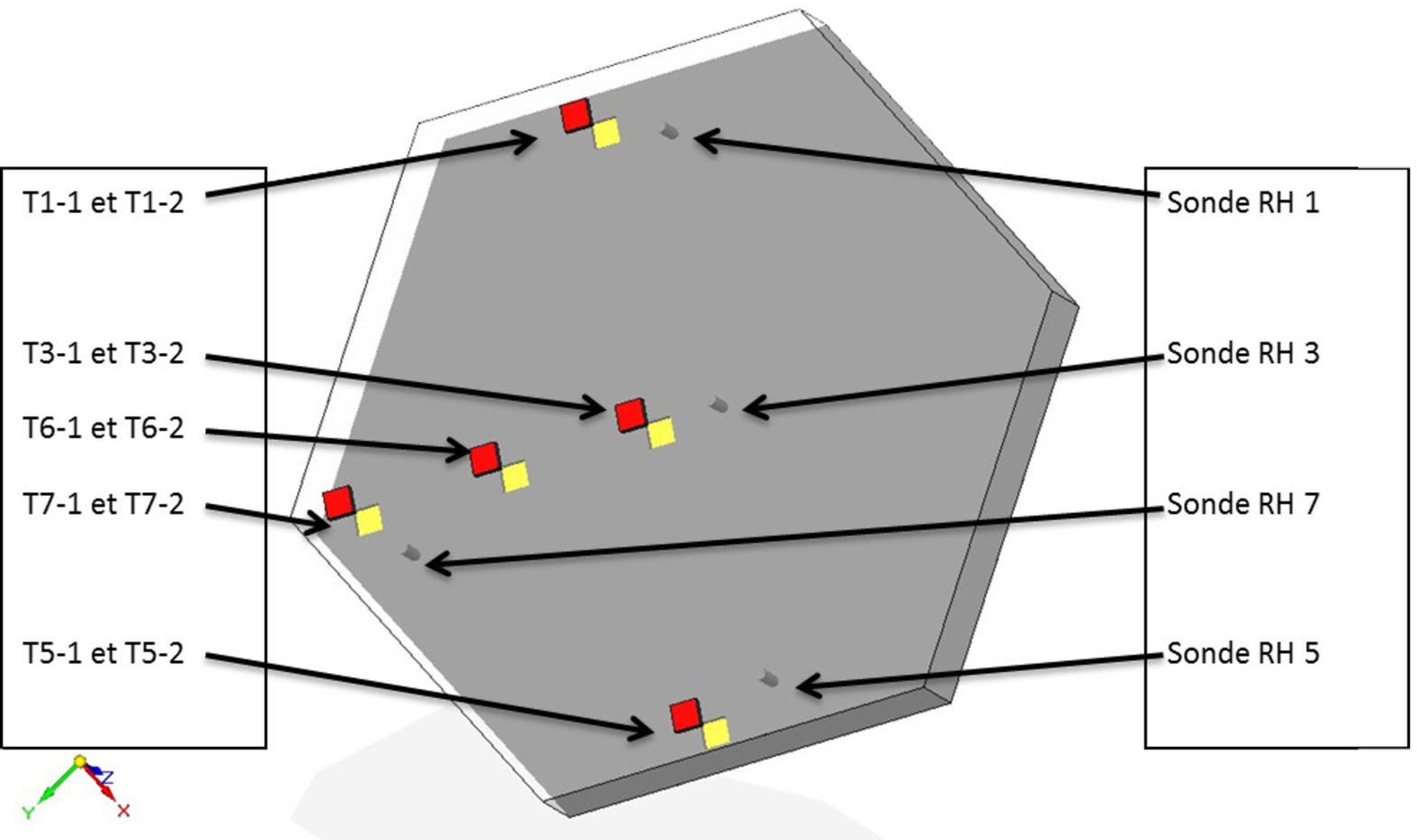}
   \end{minipage}
     \begin{minipage}[L]{7 cm}
    \includegraphics[width=5cm]{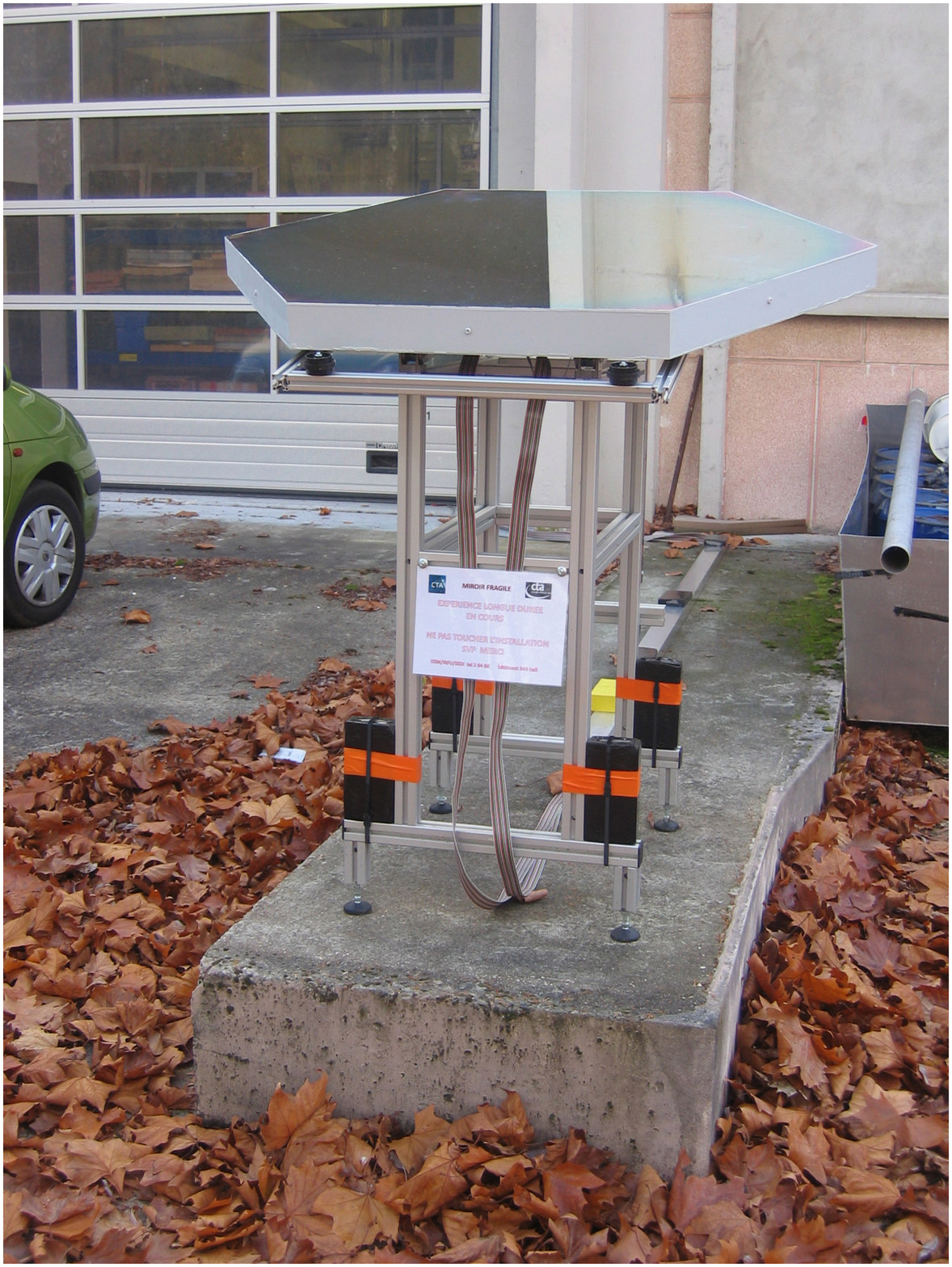}
   \end{minipage}  
   \caption{{\it Top}: Scheme of the probe positions inside the mirror. {\it Bottom}: Mirror with internal sensors placed outdoor and exposed to severe weather conditions.  \label{fig:out}} 
\end{figure}

In order to test water tightness and stability of the mirrors design under realistic weather conditions, one of the prototype mirrors has been equipped with internal probes for temperature and humidity. One set of probes was installed on the back G10 layer and a second one on the front side. This mirror was placed outdoor on Saclay site during winter time and data has been recorded for about two months, several times per day. Similar temperature and humidity probes were also installed outside the mirror. In Fig.~\ref{fig:out} are displayed the locations of the probes in the mirror, and a picture of the mirror standing outdoors. 

After this test, we verified that the amount of humidity inside the mirror stayed very small and stable even though the mirror was exposed to highly humid weather conditions. The air stayed dry inside the cells, where relative humidity was always less than 25\%  and uncorrelated with external conditions. These observations make us confidents that water does not enter the cells. Note that the mirror used for these tests had holes in the sidewalls whereas mirrors are completely sealed in the baseline design, making them even less subject to water entering. 

During the duration of the tests, the temperature difference between the two sides of the mirror (back and aluminized front) was always smaller than 2$^{\circ}$C, indicating that no stress is introduced in the materials by a large temperature gradient. 

\subsection{Temperature cycling}

\subsubsection{Small Samples}

A set of small, curved (R = 30 m) samples of 100 mm by 50 mm by 80 mm were built following the same procedure as for the full-size mirror and were submitted to a series of thermal cycles in order to study statistically the behavior of such structures after periodic weather condition changes.
Note that these samples are not at all representative of the full-size mirrors, as no sidewalls were added to the samples and the curvature was only in one dimension. This test was intended to detect potential problems related to the behavior of the glue used on the mirrors structure or with the G10 itself.

\begin{figure}[!h]
  \centering
    \includegraphics[width=0.6\textwidth]{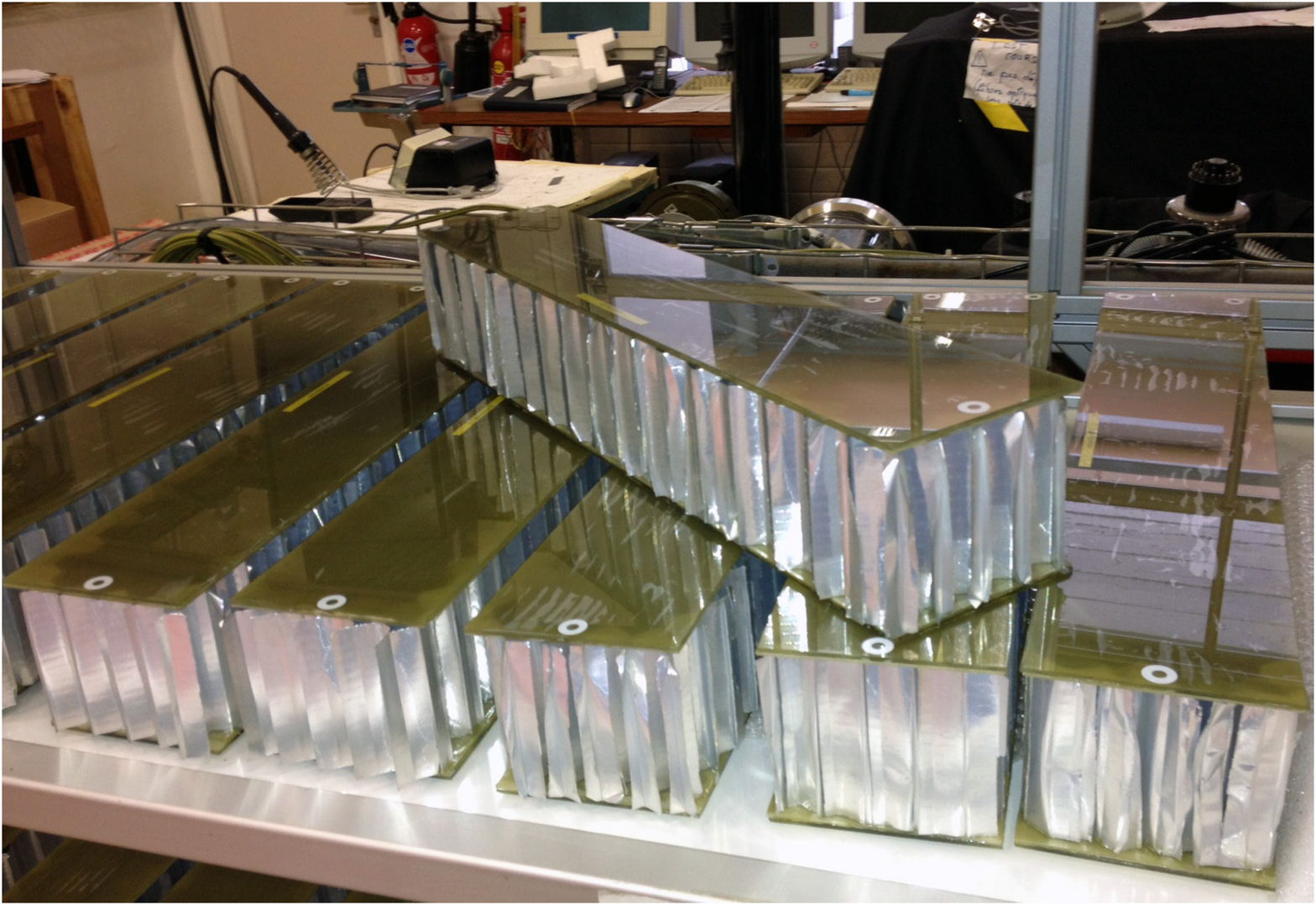}
    \includegraphics[width=0.31\textwidth]{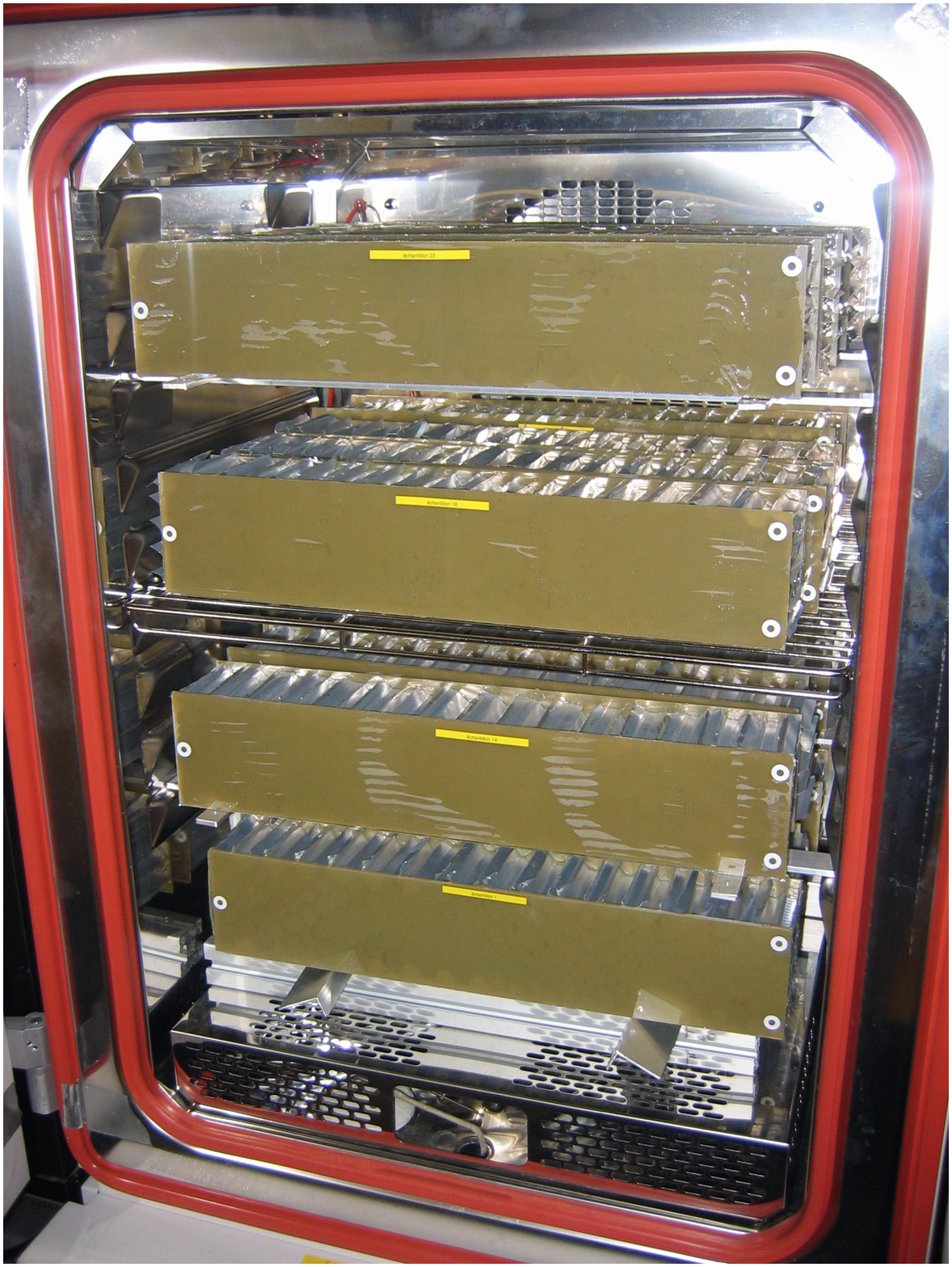}
   \caption{{\it Left}: Small samples of of 100 mm by 50 mm by 80 mm with a curvature radius of 30 m. {\it Right}: Samples inside the climate chamber. \label{fig:samples}} 
\end{figure}

A small climate chamber capable of accommodating a fraction of these samples was used (see Fig.~\ref{fig:samples}). Cycles between -20$^{\circ}$C to +60$^{\circ}$C have been performed for about 38 days, accounting for a total of 150 cycles of 6 hours each. At the end of the cycling, the whole sample survived the temperature variation without any crack or ungluing effect. Moreover, the same resistance to mechanical ripping was observed after the cycles. This indicates that the glue most likely did not degrade and it kept its adherence power.

\subsubsection{Nominal size prototype mirror}

In order to test the long-term-stability of the PSF and the durability of the materials and gluing, a prototype mirror has been submitted to temperature cycling between -25$^{\circ}$C and +60$^{\circ}$C going through different temperature gradients several times. These tests have been conducted by an external company, as no thermally-controlled room large enough is available at IRFU. The scheme followed for the cycling is shown in Tab.~\ref{tab:cyc}. The rise and fall time of the temperature in each cycle is 2 hours and the mirrors stay at constant temperature for 1 hour each time. 

\begin{table}[!h]
\centering 
	\begin{tabular}{|c|c|c|}
	\hline
	Number of cycles & Duration [hours] & Temperature range [$^{\circ}$C]  \\ \hline
 	4 &  6 & +30 to -5 \\ 
	\hline
 	4 &  6 & +30 to -10 \\ 
	\hline
	4 &  6 & +30 to -15\\ 
	\hline
 	4 &  6 & +30 to -20\\ 
	\hline
	4 &  6 & +60 to -25\\ 
	\hline	
	\end{tabular}
	\caption{Temperature cycles applied to one of the MST prototype mirrors. \label{tab:cyc}}
\end{table}

\begin{figure}[!h]
  \centering
   \begin{minipage}[T]{11 cm}
    \includegraphics[width=0.9\textwidth]{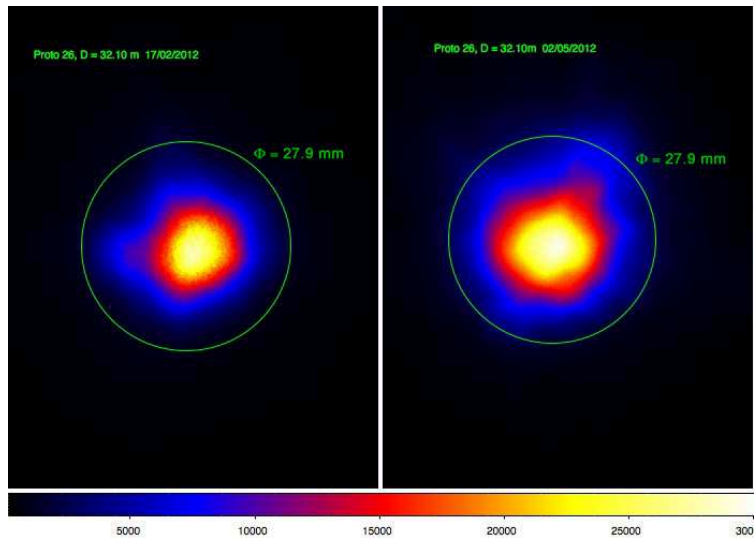}
   \end{minipage}
     \begin{minipage}[B]{11 cm}
    \includegraphics[width=0.9\textwidth]{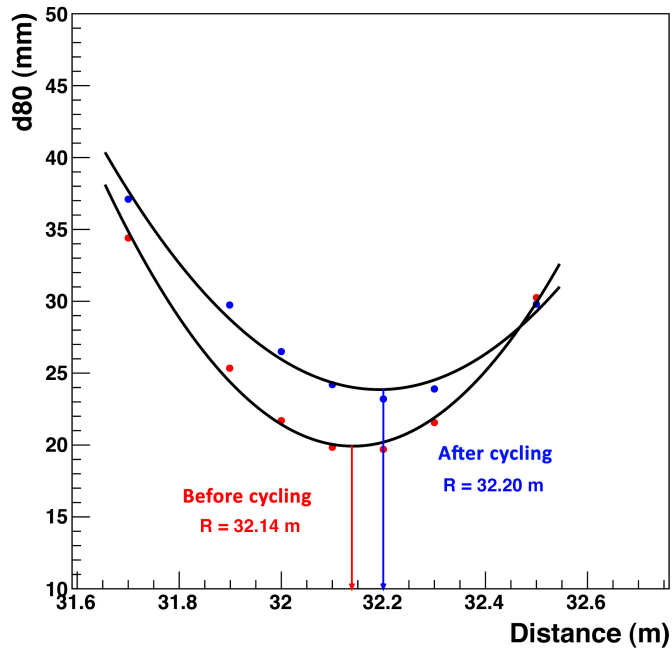}
   \end{minipage}  
   \caption{{\it Top}: Images produced at twice the focal distance of the mirror before (left) and after (right) it underwent temperature variations from -25$^{\circ}$ and +60$^{\circ}$. {\it Bottom}: Effective focal length before and after the cycling.\label{fig:temp} }
\end{figure}

The optical properties of the mirror were measured before and after the mirror underwent the temperature cycles described above. The mirror survived the test without any crack or ungluing effect. The radius of curvature of the mirror did not show significant change ($\sim$ 2$\%$) and the PSF shows a degradation of about 20$\%$, still fulfilling the CTA requirements. These results are shown on Fig.~\ref{fig:temp}.
 
Further thermal tests will be conducted in the future, in particular with the possibility of following the shift of the focal distance during the cycles themselves using a facility in DESY Zeuthen.

\subsection{Impact tests}

\begin{figure}[!h]
  \centering
    \includegraphics[width=0.5\textwidth]{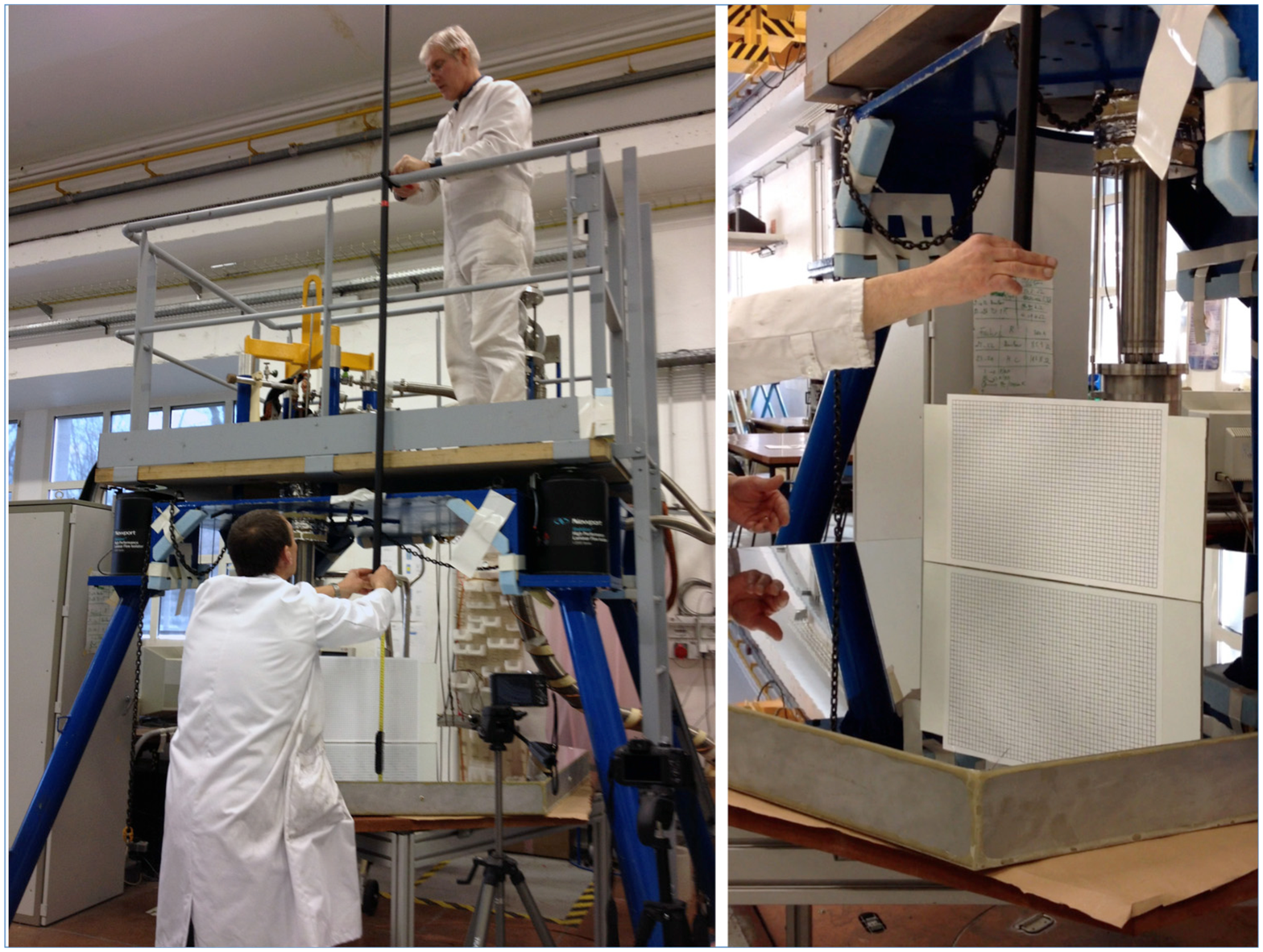}
    \includegraphics[width=0.5\textwidth]{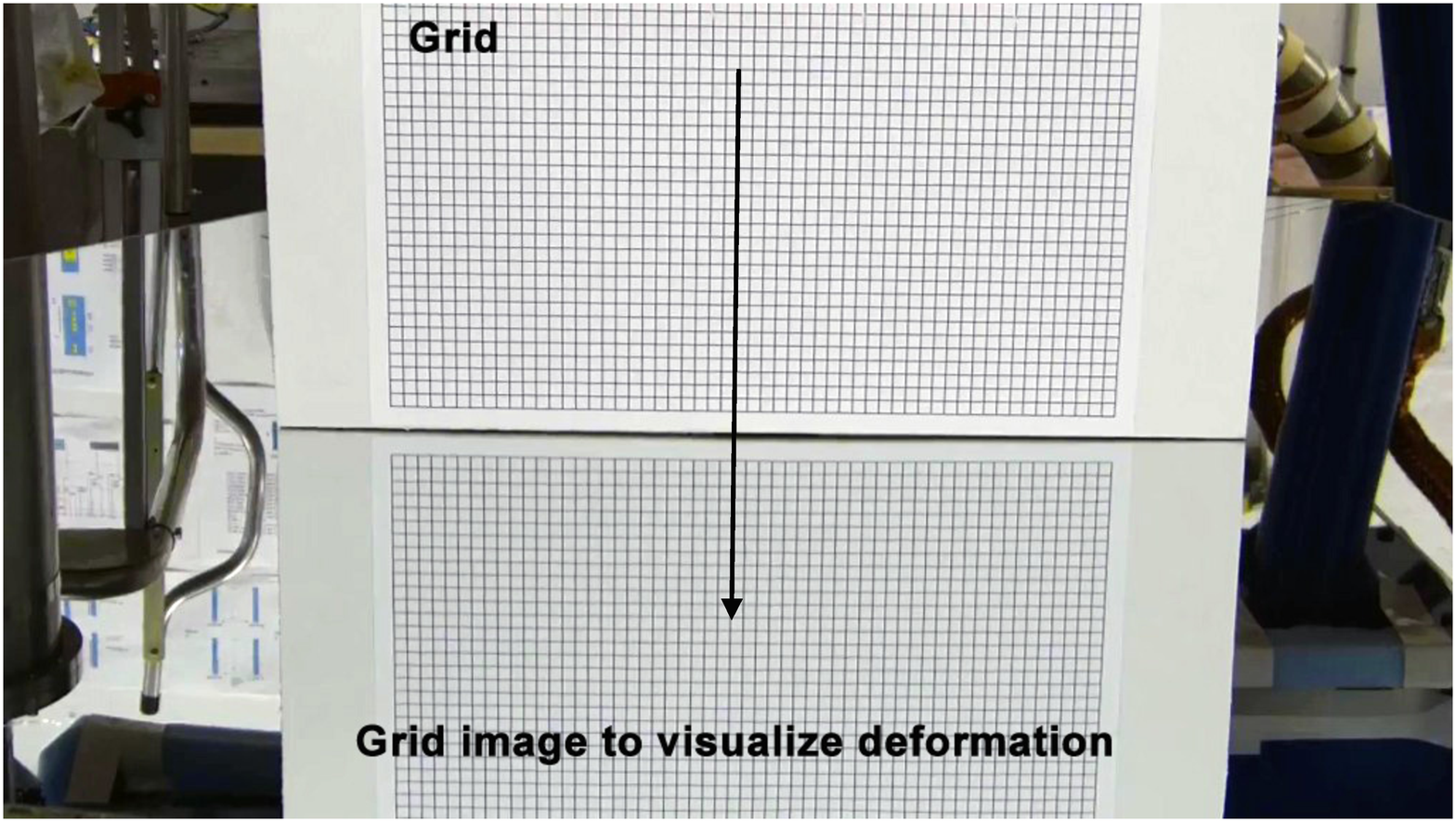}
   \caption{Impact test setup. \label{fig:impact1}}
\end{figure}

In order to perform the mechanical test simulating impacts, steel balls have been repeatedly dropped on the mirror. The mirror was placed with the coated side upwards. The ball was always released from the same place with no initial velocity. The location of the impacts was analyzed by looking at a deformation on the reflection of a grid placed nearby (see Fig.~\ref{fig:impact1}). The test was repeated a total of 10 times for each ball diameter and height.  The degrees of severity of the tests are defined by the amount of energy deposited during the impact, which depends on these two variables. Three degrees of severity have been tested during these tests (see Tab.~\ref{table3}).

\begin{table}[!h]
\centering 
	\begin{tabular}{|c|c|c|}
	\hline
	Degree of severity & Diameter [mm] & Height [cm]  \\ \hline
 	1 &  20 & 50 \\ 
	\hline
 	2 &  20 &  100 \\ 
	\hline
	3 &  30 & 50 \\ 
	\hline 	
	\end{tabular}
	 \caption{Impact tests severity levels. \label{table3}}
\end{table}
 
\begin{figure}[!h]
  \centering
    \includegraphics[width=0.4\textwidth]{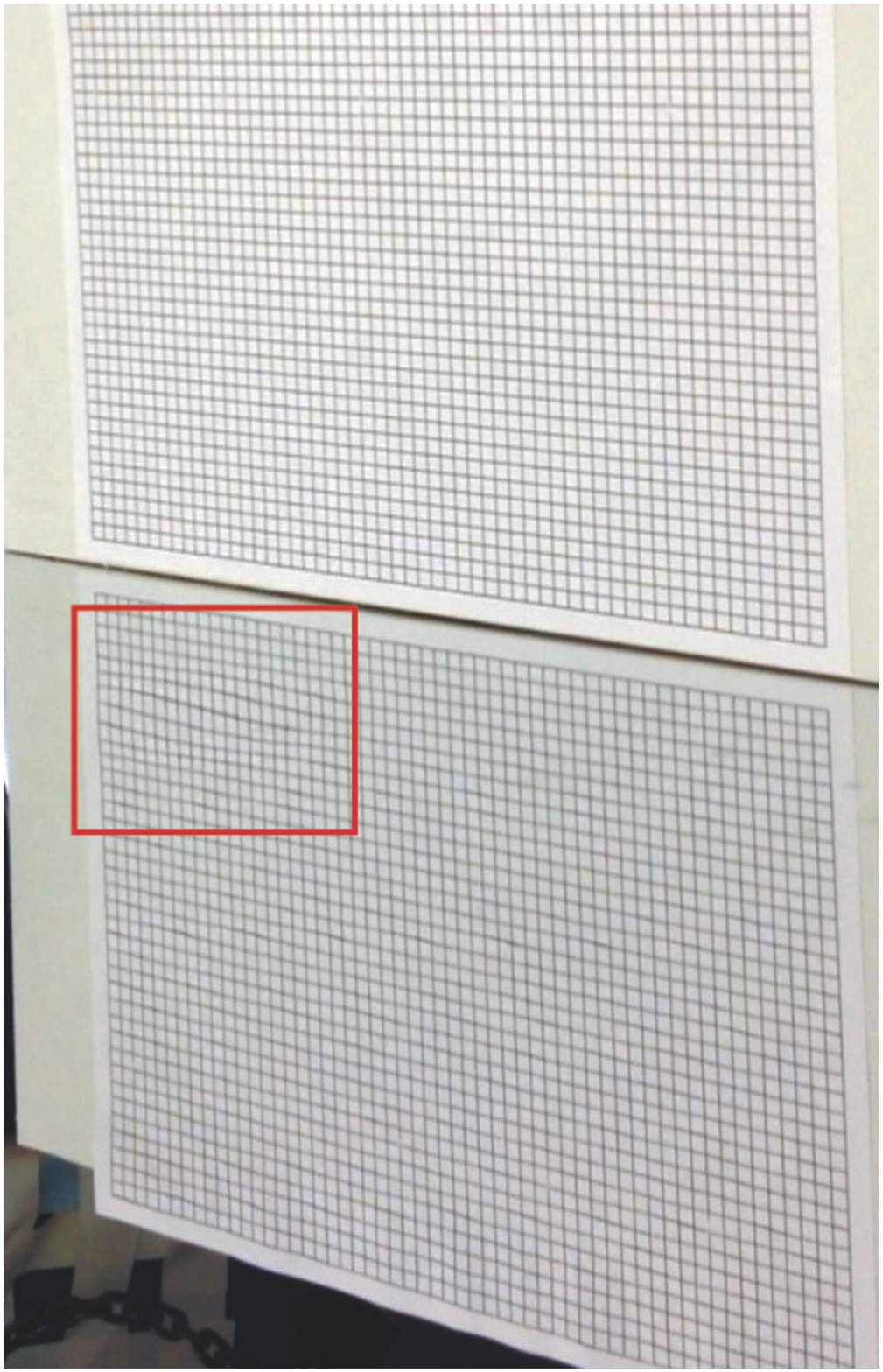}
   \caption{Deformation produced on the mirror surface by a 30 mm diameter steel ball dropped 10 times from 50 cm height. \label{fig:impact2}} 
\end{figure}

The mirror subject to these tests survived without any deformation the  severity levels 1 and 2, fulfilling the CTA specifications. Only after the level 3 severity test the mirror showed visible deformation which can be seen on the Fig.~\ref{fig:impact2}.  

\section{Conclusions and outlook}

We have presented the development of composite mirrors facets for the CTA middle-size telescopes, together with optical and mechanical tests. It is shown that the prototypes fulfill the specifications for CTA MST mirrors. They are good candidates to equip the array and  the next step is to prove the feasibility of the process in the industry, as well as its cost effectiveness. The realization of a series of 20 identical mirrors by the industry has been already launched. The analysis of the optical and mechanical properties of these mirrors will allow to assess the issue of industrialization, they will be presented in a future paper. Further tests will provide also information about long term durability of the structure itself and the reflective coating.

\section{Acknowledgements}
We would like to warmly thank our colleagues from the mirror working group in CTA, with a special thanks to Andreas F\"orster and Mos\`e Mariotti. We acknowledge the precious help from Michael Punch and Eckart Lorentz in the early stage of this work. This work benefitted a lot from the collaboration with colleagues from Erlangen OSMIN and ECAP, we would like to thank in particular Christian Faber and Friedrich Stinzing. We want to very much thank Jurgen B\"ahr for collaborating with us, as well as Vitor de Souza and Jessica Dipold. Thanks to Karl Kosack for reading the article and, at last but not least, without the dedicated and efficient effort of Gilles Decock, Jean-Luc Dominique and Thierry Chaleil this work would not have been possible. 

\bibliographystyle{elsarticle-num}
\bibliography{mirrors-bw}

\begin{thebibliography}{1}
\expandafter\ifx\csname url\endcsname\relax
  \def\url#1{\texttt{#1}}\fi
\expandafter\ifx\csname urlprefix\endcsname\relax\def\urlprefix{URL }\fi
\expandafter\ifx\csname href\endcsname\relax
  \def\href#1#2{#2} \def\path#1{#1}\fi

\bibitem{2011ExA....32..193A}
M.~{Actis}, G.~{Agnetta}, F.~{Aharonian}, A.~{Akhperjanian}, J.~{Aleksi{\'c}},
  E.~{Aliu}, D.~{Allan}, I.~{Allekotte}, F.~{Antico}, L.~A. {Antonelli},
  et~al., {Design concepts for the Cherenkov Telescope Array CTA: an advanced
  facility for ground-based high-energy gamma-ray astronomy}, Experimental
  Astronomy 32 (2011) 193--316.
\newblock \href {http://arxiv.org/abs/1008.3703} {\path{arXiv:1008.3703}},
  \href {http://dx.doi.org/10.1007/s10686-011-9247-0}
  {\path{doi:10.1007/s10686-011-9247-0}}.

\bibitem{2008SPIE.7018E..28P}
G.~{Pareschi}, E.~{Giro}, R.~{Banham}, S.~{Basso}, D.~{Bastieri},
  R.~{Canestrari}, G.~{Ceppatelli}, O.~{Citterio}, M.~{Doro}, M.~{Ghigo},
  F.~{Marioni}, M.~{Mariotti}, M.~{Salvati}, F.~{Sanvito}, D.~{Vernani}, {Glass
  mirrors by cold slumping to cover 100 m$^{2}$ of the MAGIC II Cherenkov
  telescope reflecting surface}, in: Society of Photo-Optical Instrumentation
  Engineers (SPIE) Conference Series, Vol. 7018 of Society of Photo-Optical
  Instrumentation Engineers (SPIE) Conference Series, 2008.
\newblock \href {http://dx.doi.org/10.1117/12.790404}
  {\path{doi:10.1117/12.790404}}.

\bibitem{Davies195716}
J.~M. Davies, E.~S. Cotton,
  \href{http://www.sciencedirect.com/science/article/pii/0038092X57901160}{Design
  of the quartermaster solar furnace}, Solar Energy 1~(23) (1957) 16 -- 22, the
  Proceedings of the Solar Furnace Symposium.
\newblock \href {http://dx.doi.org/10.1016/0038-092X(57)90116-0}
  {\path{doi:10.1016/0038-092X(57)90116-0}}.
\newline\urlprefix\url{http://www.sciencedirect.com/science/article/pii/0038092X57901160}

\bibitem{2011arXiv1111.2183C}
T.~{CTA Consortium}, {Contributions from the Cherenkov Telescope Array (CTA)
  Consortium to the ICRC 2011}, ArXiv e-prints (2011) pag: 26 -- 29,{\it First
  results on mirror design for CTA at Irfu-Saclay}.
\newblock \href {http://arxiv.org/abs/1111.2183} {\path{arXiv:1111.2183}}.

\end{thebibliography}

\end{document}